\documentclass{PoS}

\newcommand{\beqn}{\begin{eqnarray}}
\newcommand{\eeqn}{\end{eqnarray}}
\newcommand{\eq}[1]{(\ref{#1})}

\newcommand{\tr}{ {\rm tr} \, }

\newcommand{\fm}{\mbox{fm}}

\newcommand{\Gev}{\mbox{GeV}}

\title{$\rho$ mesons in strong abelian magnetic field in SU(3) lattice gauge theory}

\ShortTitle{$\rho$ mesons in strong abelian magnetic field in SU(3) lattice gauge theory}

\author{E.V. Luschevskaya\\
       Institute of Theoretical and Experimental Physics, Bolshaya Cheremushkinskaya 25, Moscow, 117218, Russia\\
        E-mail: \email{luschevskaya@itep.ru}}
        
                \author{O.A. Kochetkov\\
      Institute of Theoretical and Experimental Physics, Bolshaya Cheremushkinskaya 25, Moscow, 117218, Russia\\
       Institut fur Theoretische Physik, Universitat Regensburg, D-93040 Regensburg, Germany\\
        E-mail: \email{oleg.kochetkov@physik.uni-r.de}}
        
        \author{\speaker{O.V. Larina}\\
       Institute of Theoretical and Experimental Physics, Bolshaya Cheremushkinskaya 25, Moscow, 117218, Russia\\
        E-mail: \email{olarina@itep.ru}}

        \author{O.V. Teryaev\\
      Joint Institute for Nuclear Research, Dubna, Russia\\
        E-mail: \email{teryaev@theor.jinr.ru}}
        
\abstract{ 
 We  explore the masses (ground states energies) of neutral and charged $\rho$ mesons in strong   abelian magnetic field in $SU(3)$ gluodynamics.
The   energies of these particle  in the external magnetic field depends on its spin projection $s_z$ on the axis of external magnetic field. The  masses of $\rho^0$ meson with   $s_z=\pm 1$ increase with the field.  The masses  of $\rho^{\pm}$ mesons with zero spin also grow  with the magnetic field. The ground state energies  of   $\rho^{-}$ meson  with $s_z=-1$ and   $\rho^{+}$ meson  with $s_z=+1$ decrease as a function of the field, while the energies of $\rho^{+}$ meson  with $s_z=-1$ and   $\rho^{-}$ meson  with $s_z=+1$ rise with the field value.}

\FullConference{The 32nd International Symposium on Lattice Field Theory,\\
		23-28 June, 2014\\
		Columbia University New York, NY}

\begin{document}

\section{Introduction}

The exploration of hadronic matter in strong abelian magnetic field has a fundamental meaning.
The strong magnetic fields of $\sim 2\ \Gev$    may be associated with the formation of the Early Universe ~\cite{Hector:00}. 
 Now it is possible to observe such magnetic fields in non-central heavy-ion collisions in terrestrial laboratories (ALICA, RHIC, NICA, FAIR). The field value can reach up to $15 m^2_{\pi} \sim 0.27\, \Gev^2$ ~\cite{Skokov:2009}, i.e. the order of energies at which the QCD effects appear. 
 
Quantum chromodynamics in a strong magnetic field shows a lot of bright interesting effects, e.g. inverse magnetic catalysis \cite{Bruckmann:2013oba}.  Calculations on the lattices are a good method for study of the phase diagram of QCD in external magnetic field \cite{Massimo.DElia:2010,Bali:2011,Bornyakov:2013eya}.
Numerical simulations in QCD with $N_f=2$ and  $N_f=2+1$  show  that strongly interacting matter in   strong magnetic field 
posses   paramagnetic properties in the confinement and deconfinement phases \cite{Bonatti:2013,Bonatti:2014,Bali:2013}.   Equation of state for   quark-gluon plasma in   strong magnetic field was also investigated in  \cite{Levkova:2014}.
 
 In the framework of the Nambu-Jona-Lasinio model it was shown that QCD vacuum becomes a superconductor
in sufficiently strong magnetic field  ($B_c=m^2_{\rho}/e \simeq 10^{16}$ Tl)   
  \cite{Chernodub:2010,superconductivity} along the direction of the magnetic field.
 This transition to superconducting phase is accompanied
by a condensation of charged  $\rho$ mesons, their masses turns to zero.

 We have calculated the ground state energies of neutral and charged  vector  mesons as a function of external magnetic field   depending  on the spin projections  $s_z = 0,\ +1$ or $-1$ on the axis of the magnetic field in $SU(3)$ lattice gauge theory without dynamical quarks.   
 For not very large magnetic fields our data for charged particles confirm   the picture of Landau levels for pointlike particles. 
Articles  \cite{Simonov:2013} \cite{Hidaka:2012} and \cite{Liu:2014}  are also devoted to the behaviour of meson masses in the external abelian magnetic field.

 \section{Observables and details of calculations} 
 \label{Setup}

The technical details of our calculations are presented in \cite{Luschevskaya}.
We generate $SU(3)$ statistically independent
lattice gauge configurations with tadpole improved Symanzik action \cite{Bornyakov:2005}. 
Then we solve Dirac equation numerically
\begin{equation}
D \psi_k=i \lambda_k \psi_k, \  \ D=\gamma^{\mu} (\partial_{\mu}-iA_{\mu}) 
\label{Dirac}
\end{equation}
and find eigenfunctions $\psi_k$ and eigenvectors $\lambda_k$ for a test quark in the external gauge field $A_{\mu}$,  which is a sum of non-abelian $SU(3)$ gluonic field and $U(1)$ abelian constant magnetic field.

  We  add the abelian magnetic field only into Dirac operator, because our theory doesn't contain  dynamical quarks.
Our simulations has been carried out on symmetrical lattices with lattice volumes $16^6$, $18^4$ and lattice spacings $a=0.105 \fm$, $a=0.115 \fm $, $0.125\ \fm$. The number of configurations for the lattices with $a=0.115 \fm $ and $0.125\ \fm$ spacings were $200-300$,  for the finest lattice with $a=0.105\ \fm$ $95$ configurations was used.

We calculate the correlation functions in coordinate space 
\begin{equation}
\langle\psi^{\dagger}(x) O_1 \psi(x) \psi^{\dagger}(y) O_2 \psi(y)\rangle_A,
\label{observables}
\end{equation}
where  $O_1, O_2=\gamma_{\mu}, \gamma_{\nu} $ 
are Dirac gamma matrices, $\mu, \nu=1,..,4$ are Lorenz indices.
  
In order to calculate the observables \eq{observables} the quark propagators have to be computed.
The   Dirac propagator for the massive quark is
\begin{equation}
D^{-1}(x,y)=\sum_{k<M}\frac{\psi_k(x) \psi^{\dagger}_k(y)}{i \lambda_k+m}.
\label{lattice:propagator}
\end{equation}
where   $M=50$ is the number of the lowest eigenmodes.
For the correlators \eq{observables} the following equality is fulfilled
\begin{equation}
\langle \bar{\psi} O_1 \psi \bar{\psi} O_2 \psi \rangle_A=-\tr[O_1D^{-1}(x,y)O_2D^{-1}(y,x)]+\tr[O_1D^{-1}(x,x)]\tr[O_2D^{-1}(y,y)].
\label{lattice:correlator}
\end{equation}
We calculate the correlators and the  make its Fourier transformation.
For the meson ground state  we have to choose $\langle\textbf{p}\rangle=0$.
 For particles with zero momentum their energy is equal to its mass $E_0=m_0$.
The  expansion of correlation function to exponential series has the form
\begin{equation}
\tilde{C}(n_t)=\langle \psi^{\dagger}(\textbf{0},n_t) O_1 \psi(\textbf{0},n_t) \psi^{\dagger}(\textbf{0},0) O_2 \psi(\textbf{0},0)\rangle_A =
\sum_k\langle 0|O_1|k \rangle \langle k|O^{\dagger}_{2}|0 \rangle e^{-n_t a E_k},
\label{sum}
 \end{equation}
 where $a$ is the lattice spacing, $n_t$ is the number of  nodes in the time direction, $E_k$ is the energy of the state with quantum number $k$.
From expansion \eq{sum} one can see that for large $n_t$ the main contribution comes from the ground state.
Because of periodic  boundary conditions on the lattice  the main contribution to the ground state has the following form
\begin{equation}
\tilde{C}_{fit}(n_t)=A_0 e^{-n_t a  E_0} + A_0 e^{-(N_T-n_t)  a E_0}=
2A_0 e^{-N_T a E_0/2} \cosh ((N_T-n_t) a E_0),
 \label{sum33}
\end{equation}
 where  $A_0$ is a constant, $E_0$ is the energy of the ground state.
 
Mass of the ground state can be evaluated fitting the correlator \eq{sum} with \eq{sum33} function.
In order to minimize the errors and exclude the contribution of excited states we take various values of $n_t$ from the interval $5 \leq n_t \leq N_T-5$.

\section{Results}
\label{Results}

The correlators of vector currents in various spatial dimensions have the following form
\begin{equation}
C_{xx}^{VV}=\langle \bar{\psi}(\textbf{0},n_t) \gamma_1 \psi(\textbf{0},n_t)
    \bar{\psi}(\textbf{0},0) \gamma_1 \psi(\textbf{0},0)\rangle,
\end{equation}
\begin{equation}
C_{yy}^{VV}=\langle \bar{\psi}(\textbf{0},n_t) \gamma_2 \psi(\textbf{0},n_t)
    \bar{\psi}(\textbf{0},0) \gamma_2 \psi(\textbf{0},0)\rangle,
\end{equation}
  \begin{equation}
 C_{zz}^{VV}=\langle \bar{\psi}(\textbf{0},n_t) \gamma_3 \psi(\textbf{0},n_t)
    \bar{\psi}(\textbf{0},0) \gamma_3 \psi(\textbf{0},0)\rangle.
 \end{equation}
 
 The energy of the ground state or the mass of vector $\rho$ meson with $s_z= 0$ spin projection could be obtained from the $C_{zz}^{VV}$ correlator. 
 The combinations of correlators 
    \begin{equation}
C^{VV}(s_z=\pm 1)= C^{VV}_{xx}+C^{VV}_{yy} \pm i(C^{VV}_{xy}-C^{VV}_{yx}).
\label{eq:CVV1}
    \end{equation}
     correspond to the vector particles with quantum numbers $s_z=+1$ and $s_z=-1$.
     
\begin{figure}[htb]
\begin{center}
\begin{tabular}{cc}
 \includegraphics[height=3.3in, angle=-90]{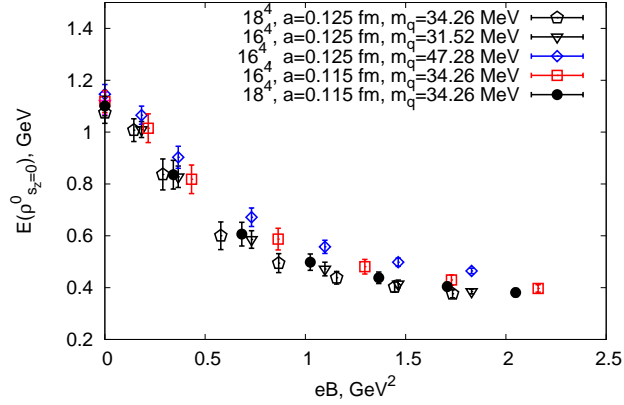}
\end{tabular}
\caption{The ground state energy  of the neutral  $\rho^0$ meson with spin   $s_z=0$  as a function 
of  magnetic field  for lattice volume $16^4$ and $18^4$, lattice spacings $a=0.115\fm$ and $0.125\ \fm$ and various bare quark masses.}
\label{fig:mrho_B2_s0}
\end{center}
\end{figure}

    In Fig.\ref{fig:mrho_B2_s0} we see the mass of the state with zero spin which was obtained from the correlator  $C_{zz}^{VV}$. We expect that at small magnetic field this state corresponds to the neutral $\rho^0$ meson with zero spin projection on the axis $z$. The mass of the state diminishes with the magnetic field at small $eB$.
 In nature at strong magnetic field the branching for the  decay $\rho^0\rightarrow \pi^0 \gamma$ have to be large.  It is not easy to  distinguish  between $\rho^0(s_z=0)$ and $\pi^0(s=0)$ on the lattice because they have the same quantum numbers. We do not make it here. 
 
\begin{figure}[htb]
\begin{center}
\begin{tabular}{cc}
 \includegraphics[height=3.3in, angle=-90]{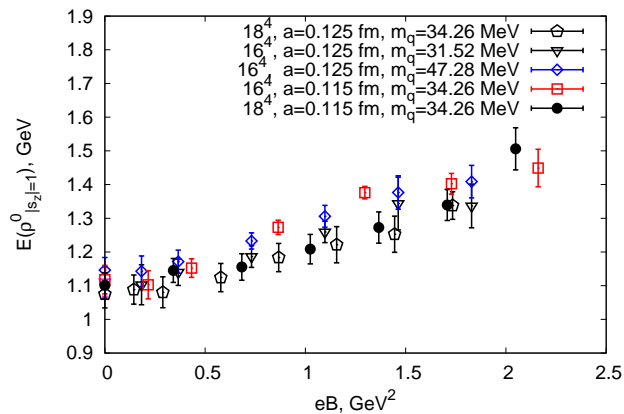}
\end{tabular}
\caption{The same that in Fig.1 but for  non-zero spin $s_z=\pm 1$  as a function of magnetic field for lattice volumes  $16^4$ and $18^4$, lattice spacings $a=0.115, \fm$ and $0.125\ \fm$ and various bare quark masses.}
\label{fig:mrho_B2_s1}
\end{center}
\end{figure}

Fig.\ref{fig:mrho_B2_s1} shows the mass of the neutral $\rho^0$ meson with spin projections $s_z=\pm 1$.
The masses for $s_z=-1$ and $s_z=+1$ grow with the field value and coincide with each other because the imaginary part of \eq{eq:CVV1} is zero.
This is a manifestation of C-parity of   $\rho^0$ meson. From Fig.\ref{fig:mrho_B2_s0} and \ref{fig:mrho_B2_s1} we  see that the data cohere  for $16^4$ and $18^4$ lattice volumes and the same lattice spacings $a=0.115$ fm. Lattice spacing effects are also not large.
 \begin{figure}[htb]
\begin{center}
\begin{tabular}{cc}
 \includegraphics[height=3.6in, angle=-90]{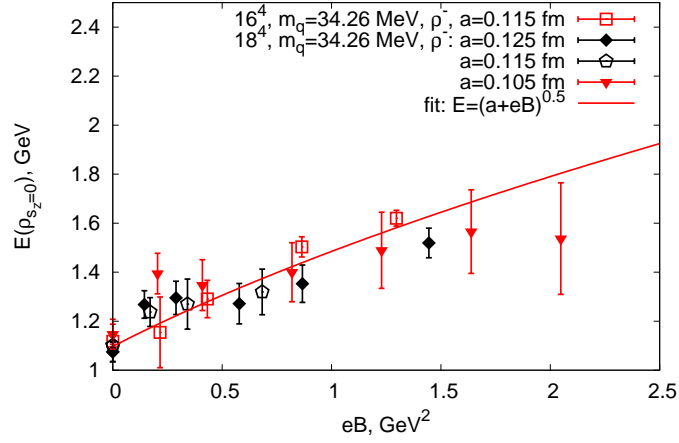}
\end{tabular}
\caption{The ground state energy  of the charged vector meson  $\rho^{-}$   with zero  spin $s_z=0$  as a function of magnetic field for    $16^4$ and $18^4$ lattices, $a=0.105,\fm$, $a=0.115,\fm$ and $0.125\ \fm$ and various bare quark masses.}
\label{fig:ch_mrho_B2_s0}
\end{center}
\end{figure}
   \begin{figure}[htb]
\begin{center}
\begin{tabular}{cc}
 \includegraphics[height=3.3in, angle=-90]{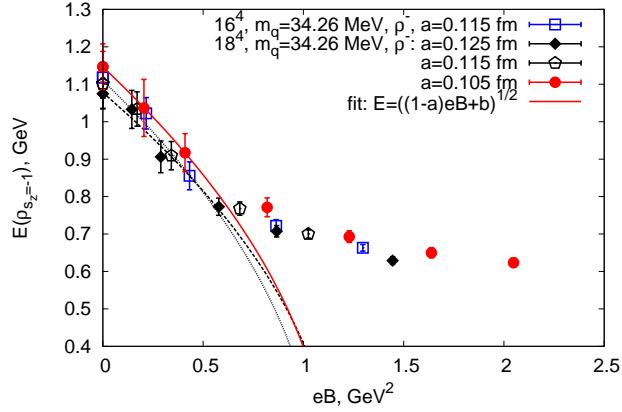}
\end{tabular}
\caption{The same that in Fig.3  but  for  $\rho^-$ meson    with   spin $s_z= -1$  versus the  field value for lattice volumes  $16^4$ and $18^4$, lattice spacings $a=0.105,\fm$, $a=0.115,\fm$ and $0.125\ \fm$ and various bare quark masses. Fits are made for the   $18^4$ lattice volume; solid line corresponds to the   $a=0.105\ \fm$ lattice spacing, dot dashed line is for $0.115\ \fm$, dashed line is for $a=0.125\ \fm$.}
\label{fig:ch_mrho_B2_s-1}
\end{center}
\end{figure}
 \begin{figure}[htb]
\begin{center}
\begin{tabular}{cc}
 \includegraphics[height=3.3in, angle=-90]{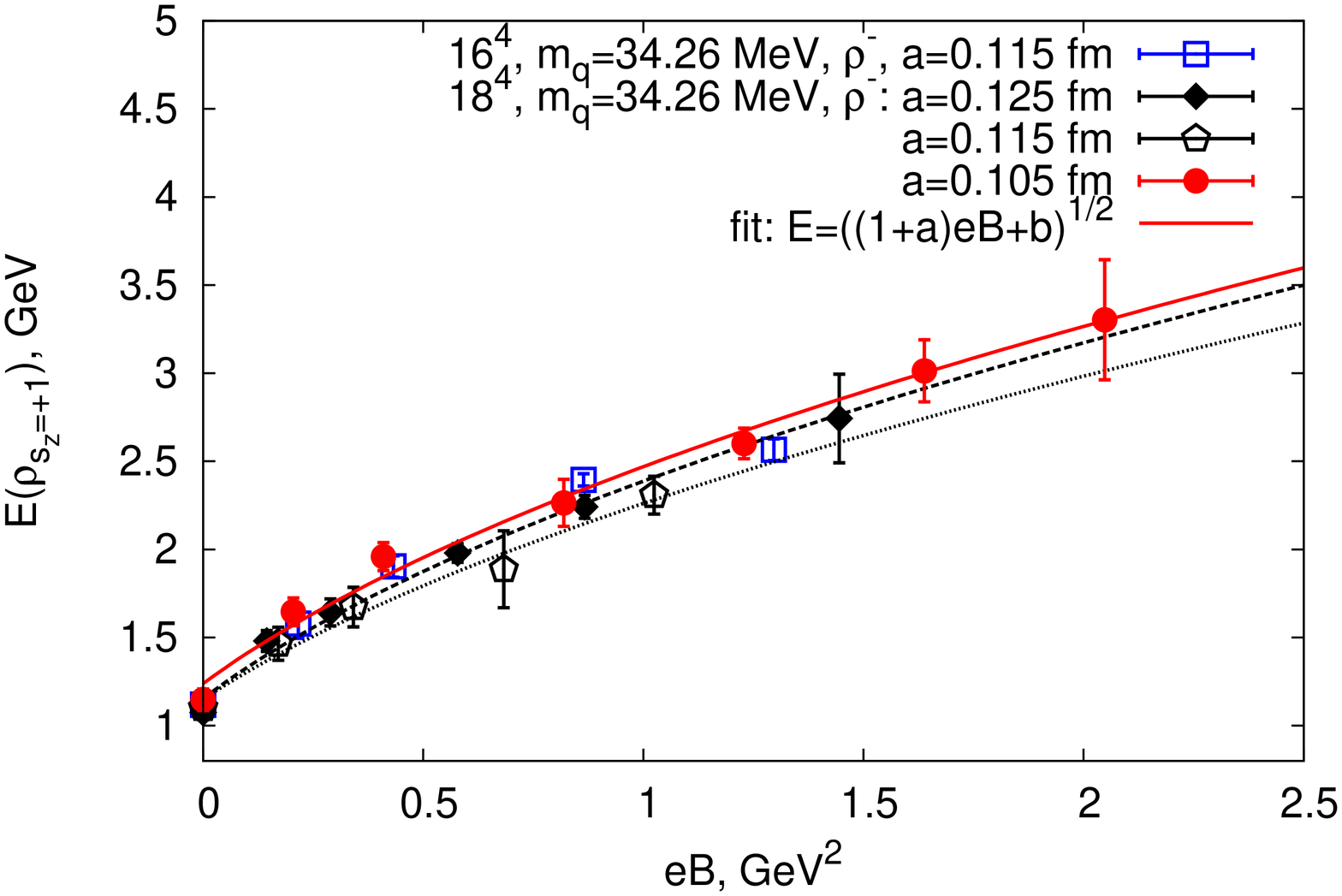}
\end{tabular}
\caption{The same  that in Fig.4 but for  spin projection $s_z= +1$  versus the magnetic field value.}
\label{fig:fig:ch_mrho_B2_s+1}
\end{center}
\end{figure}
The results are in qualitative agreement with the calculations made in the $SU(2)$ lattice gauge theory in our previous work \cite{Luschevskaya}.

The energy levels of free charged pointlike particle in a background magnetic field parallel to $z$ axis
 \begin{equation}
E^2=p^2_z+(2n+1)|qB|-gs_zqB+m^2(B=0),
\label{Landaulevels}
    \end{equation}
where $p$ is the momentum, $n$ is the number of energy level, $g$-factor characterizes magnetic properties of the particle, $q$ is the charge of the particle,
$m^2(B=0)$ is the particle mass at $B=0$. In our case $p=0$, $n=0$ and $g$ and $m^2(B=0)$ are the parameters of the fitting function. 
This equation doesn't take into account polarizabilities of the particle and  has not be valid at  very large magnetic fields. The large fields at first sight do not manifest tachyonic mode although this requires further investigation.

Fig.\ref{fig:ch_mrho_B2_s0} depicts the mass of charged vector $\rho$ meson  with $s_z=0$.
We cannot   distinguish negative and positive vector $\rho$ mesons on the lattice because the exchange of the particle charge to the opposite  is equivalent to the exchange of $B$ to $-B$ which gives the same masses for $s_z=0$. This is in accordance with  \eq{Landaulevels}.
We make a fit $E=\sqrt{a+eB}$ of the data for  $18^4$ lattice volume and $a=0.105$ fm lattice spacing, $a=m^2(B=0)$ is a fit parameter. In spite of lack of statistics the data agree with the fit.

We shows the energies of charged $\rho^{-}$ with spin projections $s_z=-1$ and $s_z=+1$ in Figs. \ref{fig:ch_mrho_B2_s-1} and \ref{fig:fig:ch_mrho_B2_s+1} correspondingly. 
The energy of the $\rho^{-}$ ground state   with $s_z=-1$ decreases with the field value. 
The data agree with fit $E=\sqrt{(1-a)eB+b}$ for $eB \in [0,0.6\ GeV^2]$, at large magnetic field field the decrease became slower. We consider this effect as the result of non-zero polarizability of charged $\rho$ meson.  
The energy of $\rho^{-}$ ground state   with $s_z=+1$ increases with the field value. The function $E=\sqrt{(1+a)eB+b}$ gives the excellent fits for the all presented data.

During calculations we are limited to a small lattice spacing   and therefore we can not explore too large values of the magnetic fields. As a result of simple estimates we obtain the  value of the magnetic field $2.9\ \Gev^2$ for lattice spacing  $a=0.115\ \fm$  and $2.5\ Gev^2$ for $a=0.125\ \fm$, when the lattice spacing effects have to become  appreciable.

\section{Conclusions}

We explore the behaviour of the masses of  vector  $\rho$  in   $SU(3)$  lattice gauge theory. 
We found that  masses of neutral vector $\rho^0$ mesons with zero and non-zero spin projection on the direction of the magnetic field differ from each other. Masses with $s_z=0$ decrease with increasing magnetic field, while the masses with $s_z=\pm 1$  increase under the same conditions. We consider this phenomenon as a result of the anisotropy produced by the strong magnetic field. We did not find condensation of neutral mesons, i.e. any indications of the existence of a phase of superfluidity in the confinement phase. 

The masses of charged $\rho$ mesons cohere with the behaviour of Landau levels at not very large magnetic fields.
We didn't observe  some evidences in favour of condensation of charged vector mesons in the considered range of fields. Condensation of charged $\rho$ mesons might indicate the existence of superconductivity in QCD at high magnetic fields. 
The existence of a superconducting phase in QCD at high values of the magnetic field B ~\cite{Chernodub:2010}  is  still a hot topic for discussions.

\section{Acknowledgments} 
 
 This work was carried out   with the financial support of Grant of President  MK-6264.2014.2 and FRRC grant of Rosatom SAEC  and Helmholtz Assotiation. The authors are grateful to  FAIR-ITEP supercomputer center where these numerical calculations were performed.

\end{document}